%Paper: hep-th/9302057
%From: Sonia Stanciu <stanciu@PIB1.physik.uni-bonn.de>
%Date: Sun, 14 Feb 1993 06:45:46 MEZ

%%%%%%%%%%%%%%%%%%%%%%%%%%%%%%%%%%%%%%%%%%%%%%%%%%%%%%%%%%%%%
%			    				    %
%   This is the Plain TeX file for			    %
%							    %
%							    %
%              On a New Supersymmetric KdV Hierarchy in	    %
%                    $2{-}d$ Quantum Supergravity 	    %
%							    %
%         by   						    %
%							    %
%               J.M. Figueroa-O'Farrill and S. Stanciu	    %
%							    %
%							    %
%							    %
%%%%%%%%%%%%%%%%%%%%%%%%%%%%%%%%%%%%%%%%%%%%%%%%%%%%%%%%%%%%%
%%%These are the macros for submission of papers to hep-th%%%
%%%The default setting is 12pt and 1 page/side but in the%%%%
%%%future it may allow people to choose also 10 pt and%%%%%%%
%%%2 pages/side.%%%%%%%%%%%%%%%%%%%%%%%%%%%%%%%%%%%%%%%%%%%%%
%%%%%%%%%%%%%%%%%%%%%%%%%%%%%%%%%%%%%%%%%%%%%%%%%%%%%%%%%%%%%
%
\def\unlockat{\catcode`\@=11}
\def\lockat{\catcode`\@=12}
\unlockat
\def\d@f@ult{} \newif\ifamsfonts \newif\ifafour
%
% \def\m@ssage{\immediate\write16}  \m@ssage{}
% \m@ssage{hep-th preprint macros.  Last modified 16/10/92 (jmf).}
% \message{These macros work with AMS Fonts 2.1 (available via ftp from}
% \message{e-math.ams.com).  If you have them simply hit "return"; if}
% \message{you don't, type "n" now: }
% \endlinechar=-1  %don't add spaces at end of line
% \read-1 to\@nswer
% \endlinechar=13
% \ifx\@nswer\d@f@ult\amsfontstrue
%     \m@ssage{(Will load AMS fonts.)}
% \else\amsfontsfalse\m@ssage{(Won't load AMS fonts.)}\fi
% %
% \message{The default papersize is A4.  If you use US 8.5" x 11"}
% \message{type an "a" now, else just hit "return": }
% \endlinechar=-1  %don't add spaces at end of line
% \read-1 to\@nswer
% \endlinechar=13
% \ifx\@nswer\d@f@ult\afourtrue
%     \m@ssage{(Using A4 paper.)}
% \else\afourfalse\m@ssage{(Using US 8.5" x 11".)}\fi
% %
% \nonstopmode
%
%%%%%%%%%%%%%%%%%%%%%%
%%%Font definitions%%%
%%%%%%%%%%%%%%%%%%%%%%
%

\font\twelverm=cmr12
\font\ninerm=cmr9
\font\sixrm=cmr6
\font\fourteenbf=cmbx12 scaled\magstep1
\font\twelvebf=cmbx12
\font\ninebf=cmbx9
\font\sixbf=cmbx6
\font\fourteeni=cmmi12 scaled\magstep1      \skewchar\fourteeni='177
\font\twelvei=cmmi12                        \skewchar\twelvei='177
\font\ninei=cmmi9                           \skewchar\ninei='177
\font\sixi=cmmi6                            \skewchar\sixi='177
\font\fourteensy=cmsy10 scaled\magstep2     \skewchar\fourteensy='60
\font\twelvesy=cmsy10 scaled\magstep1       \skewchar\twelvesy='60
\font\ninesy=cmsy9                          \skewchar\ninesy='60
\font\sixsy=cmsy6                           \skewchar\sixsy='60
\font\fourteenex=cmex10 scaled\magstep2
\font\twelveex=cmex10 scaled\magstep1

\font\ninex=cmex9
\ifamsfonts
   
   \font\sixex=cmex7 at 6pt
   
\else
   
   \font\sixex=cmex10 at 6pt
   
\fi
\font\fourteensl=cmsl10 scaled\magstep2
\font\twelvesl=cmsl10 scaled\magstep1

\font\sevensl=cmsl10 at 7pt
\font\sixsl=cmsl10 at 6pt

\font\fourteenit=cmti12 scaled\magstep1
\font\twelveit=cmti12

\font\fourteentt=cmtt12 scaled\magstep1
\font\twelvett=cmtt12
\font\fourteencp=cmcsc10 scaled\magstep2
\font\twelvecp=cmcsc10 scaled\magstep1

\ifamsfonts
   
\else
   
\fi
\newfam\cpfam
\font\fourteenss=cmss12 scaled\magstep1
\font\twelvess=cmss12
\font\tenss=cmss10
\font\niness=cmss9

\font\sevenss=cmss8 at 7pt
\font\sixss=cmss8 at 6pt
\newfam\ssfam
\newfam\msafam \newfam\msbfam \newfam\eufam
\ifamsfonts
 \font\fourteenmsa=msam10 scaled\magstep2
 \font\twelvemsa=msam10 scaled\magstep1
 \font\tenmsa=msam10
 \font\ninemsa=msam9
 \font\sevenmsa=msam7
 \font\sixmsa=msam6
 \font\fourteenmsb=msbm10 scaled\magstep2
 \font\twelvemsb=msbm10 scaled\magstep1
 \font\tenmsb=msbm10
 \font\ninemsb=msbm9
 \font\sevenmsb=msbm7
 \font\sixmsb=msbm6
 \font\fourteeneu=eufm10 scaled\magstep2
 \font\twelveeu=eufm10 scaled\magstep1
 \font\teneu=eufm10
 \font\nineeu=eufm9
 
 \font\seveneu=eufm7
 \font\sixeu=eufm6
 \def\hexnumber@#1{\ifnum#1<10 \number#1\else
  \ifnum#1=10 A\else\ifnum#1=11 B\else\ifnum#1=12 C\else
  \ifnum#1=13 D\else\ifnum#1=14 E\else\ifnum#1=15 F\fi\fi\fi\fi\fi\fi\fi}
 \def\hexmsa{\hexnumber@\msafam}
 \def\hexmsb{\hexnumber@\msbfam} 
\fi
\newdimen\b@gheight             \b@gheight=12pt
\newcount\f@ntkey               \f@ntkey=0
\def\f@m{\afterassignment\samef@nt\f@ntkey=}
\def\samef@nt{\fam=\f@ntkey \the\textfont\f@ntkey\relax}
\def\rm{\f@m0 }
\def\mit{\f@m1 }
\def\cal{\f@m2 }
\def\it{\f@m\itfam}
\def\sl{\f@m\slfam}
\def\bf{\f@m\bffam}
\def\tt{\f@m\ttfam}
\def\caps{\f@m\cpfam}
\def\ssf{\f@m\ssfam}
\ifamsfonts
 \def\msa{\f@m\msafam}
 \def\msb{\f@m\msbfam} 
 \def\eu{\f@m\eufam}
\else
  \let\eu=\bf
\fi
\def\fourteenpoint{\relax
    \textfont0=\fourteencp          \scriptfont0=\tenrm
      \scriptscriptfont0=\sevenrm
    \textfont1=\fourteeni           \scriptfont1=\teni
      \scriptscriptfont1=\seveni
    \textfont2=\fourteensy          \scriptfont2=\tensy
      \scriptscriptfont2=\sevensy
    \textfont3=\fourteenex          \scriptfont3=\twelveex
      \scriptscriptfont3=\tenex
    \textfont\itfam=\fourteenit     \scriptfont\itfam=\tenit
    \textfont\slfam=\fourteensl     \scriptfont\slfam=\tensl
      \scriptscriptfont\slfam=\sevensl
    \textfont\bffam=\fourteenbf     \scriptfont\bffam=\tenbf
      \scriptscriptfont\bffam=\sevenbf
    \textfont\ttfam=\fourteentt
    \textfont\cpfam=\fourteencp
    \textfont\ssfam=\fourteenss     \scriptfont\ssfam=\tenss
      \scriptscriptfont\ssfam=\sevenss
    \ifamsfonts
       \textfont\msafam=\fourteenmsa     \scriptfont\msafam=\tenmsa
         \scriptscriptfont\msafam=\sevenmsa
       \textfont\msbfam=\fourteenmsb     \scriptfont\msbfam=\tenmsb
         \scriptscriptfont\msbfam=\sevenmsb
       \textfont\eufam=\fourteeneu     \scriptfont\eufam=\teneu
         \scriptscriptfont\eufam=\seveneu \fi
    \samef@nt
    \b@gheight=14pt
    \setbox\strutbox=\hbox{\vrule height 0.85\b@gheight
                                depth 0.35\b@gheight width\z@ }}
\def\twelvepoint{\relax
    \textfont0=\twelverm          \scriptfont0=\ninerm
      \scriptscriptfont0=\sixrm
    \textfont1=\twelvei           \scriptfont1=\ninei
      \scriptscriptfont1=\sixi
    \textfont2=\twelvesy           \scriptfont2=\ninesy
      \scriptscriptfont2=\sixsy
    \textfont3=\twelveex          \scriptfont3=\ninex
      \scriptscriptfont3=\sixex
    \textfont\itfam=\twelveit    %\scriptfont\itfam=\nineit
    \textfont\slfam=\twelvesl    %\scriptfont\slfam=\ninesl
      \scriptscriptfont\slfam=\sixsl
    \textfont\bffam=\twelvebf     \scriptfont\bffam=\ninebf
      \scriptscriptfont\bffam=\sixbf
    \textfont\ttfam=\twelvett
    \textfont\cpfam=\twelvecp
    \textfont\ssfam=\twelvess     \scriptfont\ssfam=\niness
      \scriptscriptfont\ssfam=\sixss
    \ifamsfonts
       \textfont\msafam=\twelvemsa     \scriptfont\msafam=\ninemsa
         \scriptscriptfont\msafam=\sixmsa
       \textfont\msbfam=\twelvemsb     \scriptfont\msbfam=\ninemsb
         \scriptscriptfont\msbfam=\sixmsb
       \textfont\eufam=\twelveeu     \scriptfont\eufam=\nineeu
         \scriptscriptfont\eufam=\sixeu \fi
    \samef@nt
    \b@gheight=12pt
    \setbox\strutbox=\hbox{\vrule height 0.85\b@gheight
                                depth 0.35\b@gheight width\z@ }}
\twelvepoint
%
%%%%%%%%%%%%%%%%%
%%%Basic skips%%%
%%%%%%%%%%%%%%%%%
%
\baselineskip = 15pt plus 0.2pt minus 0.1pt %was 20pt ...
\lineskip = 1.5pt plus 0.1pt minus 0.1pt
\lineskiplimit = 1.5pt
\parskip = 6pt plus 2pt minus 1pt
\interlinepenalty=50
\interfootnotelinepenalty=5000
\predisplaypenalty=9000
\postdisplaypenalty=500
\hfuzz=1pt
\vfuzz=0.2pt
\dimen\footins=24 truecm % 8 truein in SB
\ifafour
 \hsize=16cm \vsize=22cm
\else
 \hsize=6.5in \vsize=9in
\fi
%
%%%%%%%%%%%%%%%
%%%Footnotes%%%
%%%%%%%%%%%%%%%
%
\skip\footins=\medskipamount
\newcount\fnotenumber
\def\clearfnotenumber{\fnotenumber=0} \clearfnotenumber
\def\fnote{\global\advance\fnotenumber by1 \generatefootsymbol
 \footnote{$^{\footsymbol}$}}
\def\fd@f#1 {\xdef\footsymbol{\mathchar"#1 }}
\def\generatefootsymbol{\iffrontpage\ifcase\fnotenumber
\or \fd@f 279 \or \fd@f 27A \or \fd@f 278 \or \fd@f 27B
\else  \fd@f 13F \fi
\else\xdef\footsymbol{\the\fnotenumber}\fi}
%
%%%%%%%%%%%%%%%%%%%%%%%%%%%%%
%%%Sections and Appendices%%%
%%%%%%%%%%%%%%%%%%%%%%%%%%%%%
%
\newcount\secnumber \newcount\appnumber
\def\clearappnumber{\appnumber=64} \def\clearsecnumber{\secnumber=0}
\clearsecnumber \clearappnumber
\newif\ifs@c % this is true if within a section as opposed to an appendix
\newif\ifs@cd % this is true if the article is being section'd
\s@cdtrue % this is the default
\def\unsectioned{\s@cdfalse\let\section=\subsection}
\newskip\sectionskip         \sectionskip=\medskipamount
\newskip\headskip            \headskip=8pt plus 3pt minus 3pt
\newdimen\sectionminspace    \sectionminspace=10pc
\def\Titlestyle#1{\par\begingroup \interlinepenalty=9999
     \leftskip=0.02\hsize plus 0.23\hsize minus 0.02\hsize
     \rightskip=\leftskip \parfillskip=0pt
     \advance\baselineskip by 0.5\baselineskip%this is a test...
     \hyphenpenalty=9000 \exhyphenpenalty=9000
     \tolerance=9999 \pretolerance=9000
     \spaceskip=0.333em \xspaceskip=0.5em
     \fourteenpoint
  \noindent #1\par\endgroup }
\def\titlestyle#1{\par\begingroup \interlinepenalty=9999
     \leftskip=0.02\hsize plus 0.23\hsize minus 0.02\hsize
     \rightskip=\leftskip \parfillskip=0pt
     \hyphenpenalty=9000 \exhyphenpenalty=9000
     \tolerance=9999 \pretolerance=9000
     \spaceskip=0.333em \xspaceskip=0.5em
     \fourteenpoint
   \noindent #1\par\endgroup }
\def\spacecheck#1{\dimen@=\pagegoal\advance\dimen@ by -\pagetotal
   \ifdim\dimen@<#1 \ifdim\dimen@>0pt \vfil\break \fi\fi}
\def\section#1{\cleareqnumber \s@ctrue \global\advance\secnumber by1
   \par \ifnum\the\lastpenalty=30000\else
   \penalty-200\vskip\sectionskip \spacecheck\sectionminspace\fi
   \noindent {\caps\enspace\S\the\secnumber\quad #1}\par
   \nobreak\vskip\headskip \penalty 30000 }
\def\undertext#1{\vtop{\hbox{#1}\kern 1pt \hrule}}
\def\subsection#1{\par
   \ifnum\the\lastpenalty=30000\else \penalty-100\smallskip
   \spacecheck\sectionminspace\fi
   \noindent\undertext{#1}\enspace \vadjust{\penalty5000}}

\def\appendix#1{\cleareqnumber \s@cfalse \global\advance\appnumber by1
   \par \ifnum\the\lastpenalty=30000\else
   \penalty-200\vskip\sectionskip \spacecheck\sectionminspace\fi
   \noindent {\caps\enspace Appendix \char\the\appnumber\quad #1}\par
   \nobreak\vskip\headskip \penalty 30000 }
\def\ack{\par\penalty-100\medskip \spacecheck\sectionminspace
   \line{\fourteencp\hfil ACKNOWLEDGEMENTS\hfil}%
\nobreak\vskip\headskip }
\def\refs{\begingroup \par\penalty-100\medskip \spacecheck\sectionminspace
   \line{\fourteencp\hfil REFERENCES\hfil}%
\nobreak\vskip\headskip \frenchspacing }
\def\endrefs{\par\endgroup}
%--- Note added
%
%%%%%%%%%%%%%%%%%%%%%%%%%%%%%%%%%
%%%Running heads and footlines%%%
%%%%%%%%%%%%%%%%%%%%%%%%%%%%%%%%%
%
\newif\iffrontpage \frontpagefalse
\headline={\hfil}
\footline={\iffrontpage\hfil\else \hss\twelverm
-- \folio\ --\hss \fi }
%
%%%%%%%%%%%%%%%%
%%%Title page%%%
%%%%%%%%%%%%%%%%
%
\newskip\frontpageskip \frontpageskip=12pt plus .5fil minus 2pt
\def\titlepage{\global\frontpagetrue\hrule height\z@ \relax
               \pubblock\relax }
\def\endtitlepage{\vfil\break\clearfnotenumber\frontpagefalse}
\def\title#1{\vskip\frontpageskip\Titlestyle{\caps #1}\vskip3\headskip}
\def\author#1{\vskip.5\frontpageskip\titlestyle{\caps #1}\nobreak}
\def\and{\par\kern 5pt \centerline{\sl and}}
\def\andauthor{\vskip.5\frontpageskip\centerline{and}\author}

\def\address#1{\par\kern 5pt\titlestyle{\it #1}}
\def\andaddress{\par\kern 5pt \centerline{\sl and} \address}

\def\abstract#1{\par\dimen@=\prevdepth \hrule height\z@ \prevdepth=\dimen@
   \vskip\frontpageskip\spacecheck\sectionminspace
   \centerline{\fourteencp ABSTRACT}\vskip\headskip
   {\noindent #1}}

\def\email#1{\fnote{\tentt e-mail: #1}}

%
%%%%%%%%%%%%%%%%%%%%
%%%some addresses%%%
%%%%%%%%%%%%%%%%%%%%
%

%
\def\Bonn{\address{%
   Physikalisches Institut der Universit\"at Bonn\break
  Nu{\ss}allee 12, W--5300 Bonn 1, GERMANY}}
%

%

%
%%%%%%%%%%%%%%%%
%%%References%%%
%%%%%%%%%%%%%%%%
%
\newcount\refnumber \def\clearrefnumber{\refnumber=0}  \clearrefnumber
\newwrite\R@fs                              %This opens a file .refs with
\immediate\openout\R@fs=\jobname.refs %the references in order of
                                            %appearance.
\def\closerefs{\immediate\closeout\R@fs} %close file so that TeX can read it
\def\refsout{\closerefs\refs
\unlockat
\input\jobname.refs
\lockat
\endrefs}
\def\refitem#1{\item{{\bf #1}}}%just bolds it so that \bf does not expand
\def\ifundefined#1{\expandafter\ifx\csname#1\endcsname\relax}
\def\[#1]{\ifundefined{#1R@FNO}%
\global\advance\refnumber by1%
\expandafter\xdef\csname#1R@FNO\endcsname{[\the\refnumber]}%
\immediate\write\R@fs{\noexpand\refitem{\csname#1R@FNO\endcsname}%
\noexpand\csname#1R@F\endcsname}\fi{\bf \csname#1R@FNO\endcsname}}
\def\refdef[#1]#2{\expandafter\gdef\csname#1R@F\endcsname{{#2}}}
%
%%%%%%%%%%%%%%%
%%%Equations%%%
%%%%%%%%%%%%%%%
%
\newcount\eqnumber \def\cleareqnumber{\eqnumber=0}
\newif\ifal@gn \al@gnfalse  % this is true if within an \eqalignno
\def\veqnalign#1{\al@gntrue \vbox{\eqalignno{#1}} \al@gnfalse}
\def\eqnalign#1{\al@gntrue \eqalignno{#1} \al@gnfalse}
\def\(#1){\relax%
\ifundefined{#1@Q}
 \global\advance\eqnumber by1
 \ifs@cd
  \ifs@c
   \expandafter\xdef\csname#1@Q\endcsname{{%
\noexpand\rm(\the\secnumber .\the\eqnumber)}}
  \else
   \expandafter\xdef\csname#1@Q\endcsname{{%
\noexpand\rm(\char\the\appnumber .\the\eqnumber)}}
  \fi
 \else
  \expandafter\xdef\csname#1@Q\endcsname{{\noexpand\rm(\the\eqnumber)}}
 \fi
 \ifal@gn
    & \csname#1@Q\endcsname
 \else
    \eqno \csname#1@Q\endcsname
 \fi
\else%
\csname#1@Q\endcsname\fi\global\let\@Q=\relax}
%
%%%%%%%%%%%%%%%%%
%%%Mathematica%%%
%%%%%%%%%%%%%%%%%
%
\newif\ifm@thstyle \m@thstylefalse
\def\mathstyle{\m@thstyletrue}
\def\proclaim#1#2\par{\smallbreak\begingroup%        small --> med???
\advance\baselineskip by -0.25\baselineskip%
\advance\belowdisplayskip by -0.35\belowdisplayskip%
\advance\abovedisplayskip by -0.35\abovedisplayskip%
    \noindent{\caps#1.\enspace}{#2}\par\endgroup%
\smallbreak}%--- defs, thms, ...                     small --> med???
\def\m@kem@th<#1>#2#3{%
\ifm@thstyle \global\advance\eqnumber by1
 \ifs@cd
  \ifs@c
   \expandafter\xdef\csname#1\endcsname{{%
\noexpand #2\ \the\secnumber .\the\eqnumber}}
  \else
   \expandafter\xdef\csname#1\endcsname{{%
\noexpand #2\ \char\the\appnumber .\the\eqnumber}}
  \fi
 \else
  \expandafter\xdef\csname#1\endcsname{{\noexpand #2\ \the\eqnumber}}
 \fi
 \proclaim{\csname#1\endcsname}{#3}
\else
 \proclaim{#2}{#3}
\fi}
\def\Thm<#1>#2{\m@kem@th<#1M@TH>{Theorem}{\sl#2}}%--- Theorem
\def\Prop<#1>#2{\m@kem@th<#1M@TH>{Proposition}{\sl#2}}%--- Proposition
\def\Def<#1>#2{\m@kem@th<#1M@TH>{Definition}{\rm#2}}%--- Definition
\def\Lem<#1>#2{\m@kem@th<#1M@TH>{Lemma}{\sl#2}}%--- Lemma
\def\Cor<#1>#2{\m@kem@th<#1M@TH>{Corollary}{\sl#2}}%--- Corollary
\def\Conj<#1>#2{\m@kem@th<#1M@TH>{Conjecture}{\sl#2}}%--- Conjecture
\def\Rmk<#1>#2{\m@kem@th<#1M@TH>{Remark}{\rm#2}}%--- Remark
\def\Exm<#1>#2{\m@kem@th<#1M@TH>{Example}{\rm#2}}%--- Example
\def\Qry<#1>#2{\m@kem@th<#1M@TH>{Query}{\it#2}}%--- Query
%
%--- Proof
%
\let\Pf=\Proof

\def\<#1>{\csname#1M@TH\endcsname}
%
%%%%%%%%%%%%%%%%%%%
%%%Abbreviations%%%
%%%%%%%%%%%%%%%%%%%
%
\def\ref#1{{\bf [#1]}}%--- [ref]
%--- et al.
%--- i.e.
\def\eg{{\it e.g.\/}}%--- e.g.
%--- Cf.
\def\cf{{\it cf.\ }}%--- cf.
 %--- double left quote
\def\th{{\rm th}}%--- th as in fifth
\def\nl{\hfil\break}%--- new line
%--- 1/2
%
%%%%%%%%%%%%%%%%%
%%%Mathematics%%%
%%%%%%%%%%%%%%%%%
%
%--- def over =
\def\qed{\vrule width 0.7em height 0.6em depth 0.2em}%--- Halmos Q.E.D.
\def\QED{\enspace\qed}
%--- implies
%--- is implied by
%--- if and only if
\def\lapprox{\hbox{\lower3pt\hbox{$\buildrel<\over\sim$}}}% approx lt
\def\gapprox{\hbox{\lower3pt\hbox{$\buildrel<\over\sim$}}}% approx gt
\def\quotient#1#2{#1/\lower0pt\hbox{${#2}$}}%--- factor objects
\ifamsfonts
 \mathchardef\empty="0\hexmsb3F %--- better empty set than \emptyset
 \mathchardef\lsemidir="2\hexmsb6E % semidirect |x
 \mathchardef\rsemidir="2\hexmsb6F % semidirect x|
\else
 \let\empty=\emptyset
 \def\lsemidir{\mathbin{\hbox{\hskip2pt\vrule height 5.7pt depth -.3pt
    width .25pt\hskip-2pt$\times$}}}
 \def\rsemidir{\mathbin{\hbox{$\times$\hskip-2pt\vrule height 5.7pt
    depth -.3pt width .25pt\hskip2pt}}}
\fi
%
%--- injective map
%--- surjective map
%--- bijective map
%--- mapping
%--- long mapping
%--- isom over -->
%--- just an abbrev.
%

%
 %--- commutative diagram macro
 %--- map in complex
%
 %--- reals
 %--- complex nos.
 %--- quaternions
 %--- integers
 %--- rationals
 %--- naturals
 %--- ground field
%
%--- Hom(omorphisms)
%--- tr(ace)
\def\Tr{\mathop{\rm Tr}}%--- Tr(ace)
%--- End(omorphisms)
%--- Mor(phisms)
%--- Aut(omorphisms)
%--- aut(omorphisms)
%--- supertrace
%--- superdeterminant
%--- kernel
%--- cokernel
%--- image
%
\def\underrightarrow#1{\vtop{\ialign{##\crcr
      $\hfil\displaystyle{#1}\hfil$\crcr
      \noalign{\kern-\p@\nointerlineskip}
      \rightarrowfill\crcr}}} %--- modification of \overrightarrow
\def\underleftarrow#1{\vtop{\ialign{##\crcr
      $\hfil\displaystyle{#1}\hfil$\crcr
      \noalign{\kern-\p@\nointerlineskip}
      \leftarrowfill\crcr}}}  %--- modification of \overleftarrow

\def\comm#1#2{\left[#1\, ,\,#2\right]}%--- [ , ]
%--- { , }
%--- [ , }
%
%--- Lie derivative
\def\vder#1#2{{{{\delta}{#1}}\over{{\delta}{#2}}}}%--- vartnl derivative
\def\pder#1#2{{{\partial #1}\over{\partial #2}}}%--- partial derivative
%--- full derivative
%
%%%%%%%%%%%%%%
%%%Journals%%%
%%%%%%%%%%%%%%
%

\def\CMP#1#2#3{{\sl Comm. Math. Phys.} {\bf #1} (#2) #3}

\def\PLA#1#2#3{{\sl Phys. Lett.} {\bf #1A} (#2) #3}
\def\PLB#1#2#3{{\sl Phys. Lett.} {\bf #1B} (#2) #3}
\def\JMP#1#2#3{{\sl J. Math. Phys.} {\bf #1} (#2) #3}

\def\LMP#1#2#3{{\sl Letters in Math. Phys.} {\bf #1} (#2) #3}

\def\RMaP#1#2#3{{\sl Reports on Math. Phys.} {\bf #1} (#2) #3}

\def\MPLA#1#2#3{{\sl Mod. Phys. Lett.} {\bf A#1} (#2) #3}

\def\JDG#1#2#3{{\sl J. Diff. Geometry} {\bf #1} (#2) #3}

\lockat
%
%   These are the local macros for (SKdV-B)
%

\let\d=\partial

\def\fr#1/#2{\mathord{\hbox{${#1}\over{#2}$}}} %fractions
\def\D#1T{\ifcase#1 T \or T'\or T'' \or T''' \else T^{[#1]}\fi}

\def\comm[#1,#2]{\left[#1\,,\,#2\right]}%--- [ , ]
\def\sint{\int\nolimits_B\,}
\refdef[SuperMM]{L. Alvarez-Gaum\'e and J.L. Ma\~nes,
\MPLA{6}{1991}{2039}.}
\refdef[Luis]{L. Alvarez-Gaum\'e, H. Itoyama, J.L. Ma\~nes and A.
Zadra, {\sl Superloop equations and two-dimensional supergravity},
{\tt hep-th/9112018}.}
\refdef[Luisetal]{L. Alvarez-Gaum\'e, K. Becker, M. Becker, R.
Empar\'an, and J.L. Ma\~nes, {\sl Double scaling limit of the
superVirasoro constraints}, {\tt hep-th/9207096}.}
\refdef[Beckers]{K. Becker and M. Becker, {\sl Nonperturbative
solutions of the superVirasoro constraints}, {\tt hep-th/9301017}.}
\refdef[EarlyKdV]{M.D. Kruskal, R.M. Miura, C.S. Gardner, and N.J.
Zabusky, \JMP{11}{1970}{952}.}
\refdef[FR]{J.M. Figueroa-O'Farrill and E. Ramos,
\CMP{145}{1992}{43}.}
\refdef[Dickey]{L.~A.~Dickey,  {\sl Soliton equations and Hamiltonian
systems}, Advanced Series in Mathematical Physics Vol.12,  World
Scientific Publishing Company.}
\refdef[Confused]{J. Barcelos-Neto and A. Das, \JMP{33}{1992}{2743}.}
\refdef[Painleve]{P. Mathieu, \PLA{128}{1988}{169}.}
\refdef[SKPtwo]{J.M. Figueroa-O'Farrill, J. Mas, and E. Ramos,
\RMaP{3}{1991}{479}.}
\refdef[Oddbracks]{B.A. Kupershmidt, \LMP{9}{1985}{323}.}
\refdef[Kupskdv]{B.A. Kupershmidt, \PLA{102}{1984}{213}.}
\refdef[MaRa]{Yu.I. Manin and A.O. Radul, \CMP{98}{1985}{65}.}
\refdef[MuRa]{M. Mulase, \JDG{34}{1991}{651};\nl J.M. Rabin,
\CMP{137}{1991}{533}.}
\refdef[slax]{J.M. Figueroa-O'Farrill and E. Ramos,
\PLB{262}{1991}{265}.}
\refdef[Bouss]{J.M. Figueroa-O'Farrill and S. Stanciu, {\sl New
supersymmetrizations of the generalized KdV hierarchies}, preprint in
preparation.}
\overfullrule=0pt
\unsectioned
\mathstyle
\def\pubblock{ \line{\hfil\rm BONN--HE--93--007}
               \line{\hfil\tt hep-th/9302057}
               \line{\hfil\rm February 1993}}
\titlepage
\title{On a New Supersymmetric KdV Hierarchy in $2{-}d$ Quantum
Supergravity}
\author{Jos\'e~M.~Figueroa-O'Farrill
\email{jmf@avzw01.physik.uni-bonn.de}}
\andauthor{Sonia Stanciu\email{sonia@avzw01.physik.uni-bonn.de}}
\Bonn
\abstract{Recently a new supersymmetric extension of the KdV hierarchy
has appeared in a matrix-model-inspired approach to $2{-}d$
quantum supergravity.  Here we prove that this hierarchy is
essentially the KdV hierarchy, where the KdV field is now replaced by
an even superfield.  This allows us to find the conserved charges and
the bihamiltonian structure, and to prove its integrability.  We also
extend the hierarchy by odd flows in a supersymmetric fashion.}
\endtitlepage

\subsection{Introduction}

One of the most pleasant surprises that noncritical string theory so
far had in store for us, is its relation with classical integrable
hierarchies of the KP type.  For example, the KdV hierarchy appeared
unsuspectedly in the double scaling limit of the one-matrix model---a
fact which recurs in the multimatrix models for the generalized KdV
hierarchies, and which allows one to exactly compute correlation
functions on arbitrary topology.  This success notwithstanding, the
generalization of these techniques to the supersymmetric case is still
an open problem and the precise relation, if any, with supersymmetric
integrable hierarchies remains elusive.

In order to circumvent the problems encountered in an earlier
unsuccessful attempt (\[SuperMM]) to define a theory of noncritical
superstrings using supermatrices, a model was proposed in \[Luis] in
which one does away with the matrices all together, and takes as a
starting point the integral over the would-be eigenvalues.  By
imposing superVirasoro constraints---in analogy with the Virasoro
constraints in the one-matrix model---correlation functions and
critical exponents were calculated to first order in the topological
expansion.  Remarkably, they were found to coincide with those of
certain superconformal matter coupled to $2{-}d$ supergravity.
Recently, in \[Beckers], the model was solved for arbitrary genus and,
in the double scaling limit, a new supersymmetric extension of the KdV
hierarchy appeared.

It is the purpose of this note to identify the hierarchy and mention
some of its immediate properties: conserved charges, bihamiltonian
structure, integrability,...  The punch line is that the hierarchy is
simply a supersymmetric covariantization of the KdV hierarchy, and as
such not very different from it.  In this sense, the model solved in
\[Beckers] seems once again not to be too distantly related to the
bosonic one-matrix model, a fact already remarked in \[Luisetal] as
well as in \[Beckers].

We should remark, parenthetically, that the existence of this new
supersymmetric extension of KdV does not contradict the Painlev\'e
analysis of \[Painleve] which suggested that the only two integrable
fermionic extensions of the KdV equation are the nonsupersymmetric
extension of \[Kupskdv] and the one of \[MaRa].  This is due to the
fact that the hierarchy we study in this paper does not fit the Ansatz
in \[Painleve].  This hierarchy contains the KdV hierarchy as a
subhierarchy and not as a reduction after setting the fermionic field
to zero.

\section{The SKdV-B hierarchy}

The hierarchy of \[Beckers] is a hierarchy of flows on two variables
$u$ and $\tau$---the ``body'' of the two-point function of the
puncture operator and the first fermionic scaling variable,
respectively; although their physical interpretation is of no
relevance to our discussion.  We introduce an infinite number of even
times $\{t_0,t_1,t_2,\ldots\}$ and an infinite number of odd times
$\{\tau_0,\tau_1,\tau_2,\ldots\}$.  On $u$ the odd flows are trivial
$$
\pder{u}{\tau_k}=0\quad\forall k~,\()
$$
whereas the even flows are those of the KdV hierarchy:
$$
\veqnalign{\pder{u}{t_k} &= R_{k+1}'\(gdone)\cr
    &= \left[ \kappa^2 \d^3 + 2 u \d + 2 \d u\right] \cdot
R_k~,\(gdtwo)\cr}
$$
where the Gel'fand--Dickey polynomials $R_k=R_k(u)$ are the gradients
of the conserved charges of the KdV hierarchy and $\kappa$ is the
renormalized string coupling constant.  The equality of \(gdone)
and \(gdtwo) imply the celebrated Lenard relations between the
$R_k$, which can be translated into a recursion relation for the
flows:
$$
\pder{u}{t_{n+1}} = \left[ \kappa^2 \d^2 + 2 u + 2 \d u \d^{-1}
\right]\cdot \pder{u}{t_{n}}~.\(kdvrec)
$$
Normalizing $R_0=\fr1/2$, we can compute all the other $R_k$
recursively: $R_1 = u$, $R_2 = \kappa^2 u'' + 3 u^2$, {\it ad
nauseam\/}.  In terms of the $R_k$, the commutativity of the KdV flows
translates into
$$
\pder{R_k'}{t_{n}} = \pder{R_{n+1}'}{t_{k-1}}~,\(flowscommute)
$$
an identity that, as we will see shortly, implies the invariance of
the even flows under supersymmetry.  From the analysis in \[Beckers],
$\tau$ is given by
$$
\tau = -\sum_{k\geq 0} \tau_k R_k~,\(tau)
$$
wherefrom we can read how it evolves along the flows
$$
\pder{\tau}{\tau_k} = - R_k\quad\hbox{and}\quad
\pder{\tau}{t_n} = -\sum_{k\geq 0} \tau_k \pder{R_k}{t_n}~.\()
$$
The first nontrivial even flows were found in \[Beckers] to be
$$
\pder{u}{t_1} = \kappa^2 u''' + 6 u u' \quad\hbox{and}\quad
\pder{\tau}{t_1} = \kappa^2 \tau''' + 6 u\tau'~,\(tone)
$$
whereas the odd flows were found to be
$$
\pder{u}{\tau_1} = 0 \quad\hbox{and}\quad \pder{\tau}{\tau_1} =
-u~.\(tauone)
$$
Notice that the the first equation in \(tone) is nothing but the KdV
equation for $u$.

It was moreover observed in \[Beckers] that \(tone) is invariant under
the (global) supersymmetric transformations
$$
\delta u=\tau'\quad\hbox{\rm and}\quad \delta\tau= u~.\(SUSY)
$$
In fact, as we will show in a moment, this continues to be the case
for all the even flows.  On the other hand, the odd flows are not
supersymmetric,  for whereas $\tau$ evolves, its supersymmetric
partner $u$ does not.  Nevertheless, one can modify the odd flows to
make them supersymmetric.  We will comment on this further on.

\Prop<FlowsComm>{The even flows are invariant under \(SUSY),
while the odd flows satisfy
$$
\comm[\delta,\pder{}{\tau_n}]=-\pder{}{t_{n-1}}~.\(oddflows)
$$
}
\Pf We first consider the even flows:
$$
\eqnalign{\comm[\delta,\pder{}{t_n}] u
          &=\left(\vder{R_{n+1}}{u}\cdot\tau'\right)' -
\pder{\tau'}{t_n}\cr
          &=-\left(\sum_{k\geq0} \tau_k\pder{R_{n+1}'}{t_{k-1}}\right)'
+ \left(\sum_{k\geq 0}\tau_k\pder{R_k'}{t_n}\right)' = 0~,\cr}
$$
where we have used \(flowscommute).  Notice that this allows us to
rewrite the flows on $\tau$ in a simpler way:
$$
\pder{\tau}{t_n}=\delta R_{n+1}~.\(flowsontau)
$$
{}From this, the analog result for $\tau$ follows trivially, because
$$
\delta\pder{\tau}{t_n}= R_{n+1}' =\pder{u}{t_n} =
\pder{}{t_n}\delta\tau~.\()
$$
On the other hand, for the odd flows we obtain for $u$
$$
\comm[\delta,\pder{ }{\tau_n}]u=\pder{\tau'}{\tau_n} =
-\pder{u}{t_{n-1}}~,\()
$$
whereas for $\tau$ one has
$$
\comm[\delta,\pder{ }{\tau_n}]\tau = -\delta R_n =
-\pder{\tau}{t_{n-1}}~,\()
$$
where we have once again used \(flowsontau). \QED

In summary, the subhierarchy defined by the even flows is a
supersymmetric extension of the KdV hierarchy, to which we refer in
what follows as the SKdV-B hierarchy.

\section{The SKdV-B Hierarchy is the KdV Hierarchy}

We now begin the analysis of this hierarchy.  We will find it
convenient to employ superfields in order to preserve manifest
supersymmetry and we also use freely the formalism of the formal
calculus of variations and of pseudodifferential operators, for which
we refer the reader to Dickey's book \[Dickey], and, for instance, to
\[FR] for the supersymmetric case.

Since the SKdV-B hierarchy is supersymmetric, one can express its
flows in a way that makes this manifest, whereto we introduce the
superfield $T=\tau +\theta u$, a function in a $(1|1)$ superspace.  In
superspace, the supersymmetry algebra is realized as
supertranslations, which on superfields look like $\delta T= Q T $,
where $Q=\pder{ }{\theta}-\theta \partial$.  We will denote by $D$ the
supercovariant derivative $D=\pder{ }{\theta}+\theta\partial$~, which
anticommutes with $Q$.  One can recover the fields $u$ and $\tau$ by
taking the appropriate projections: $u=DT|_{\theta =0}$ , $\tau
=T|_{\theta =0}$.

Rewriting both equations in \(tone) as a single equation on the
superfield $T$, we find\fnote{Please note that on a superfield $'$
denotes derivative with respect to $D$, whereas on components it
denotes derivative with respect to $\d$.  This should cause no
confusion.}
$$
\pder{T}{t_1}=\kappa^2 \D6T +6T'T''~,\()
$$
where ${}^{[~]}$ denotes differentiation with respect to $D$. Now
notice that if we differentiate both sides of the equation once more
with respect to $D$, we get
$$
\pder{T'}{t_1}=\kappa^2 \D7T +6T'T'''~,\()
$$
which is nothing but the KdV equation (\cf the first equation in
\(tone)) for the superfield $T'=u+\theta\tau'$.  In fact, as we now
show, this continues to be the case for all the other equations of the
hierarchy; whence we will be able to conclude that the SKdV-B
hierarchy is essentially equivalent to the KdV hierarchy.

This may require some explanation.  The abstract KdV hierarchy is
defined as the hierarchy of isospectral deformations of the Lax
operator $L= \kappa^2 \d^2 + u$, where $u$ is simply a commuting
variable generating a differential ring.  Particular representations
of this abstract KdV hierarchy are obtained by letting $u$ be, for
instance, a smooth function on the circle or a rapidly decaying smooth
function on the real line.  A more exotic representation can be
defined by taking $u$ to be an even superfield, \eg, $T'$.  We claim
that the hierarchy so obtained is precisely SKdV-B.  For notational
convenience we will denote by KdV($T'$) the KdV hierarchy with $T'$ as
the basic variable, and reserve KdV for when the basic variable is
$u$.

Consecutive flows in both the KdV($T'$) and SKdV-B hierarchies are
related by a recursion relation.  This means that knowing the first
flow one can obtain all the others by repeated application of a
recursion operator.  We have seen that the first flows of both
hierarchies agree, thus all we need to show in order to prove the
equivalence is that the recursion operators are the same.

The recursion relation for the flows of the KdV($T'$) hierarchy can be
read off from \(kdvrec) and is given by
$$
\pder{T'}{t_{n+1}}=\left[\kappa^2\partial^2 +2 \partial
T'\partial^{-1} + 2T'\right]\cdot \pder{T'}{t_n}~.\()
$$
Stripping off a $D$ from both sides, we can rewrite this as
$$
\pder{T}{t_{n+1}}=\left[\kappa^2\partial^2 +2 D T'D^{-1} + 2
D^{-1}T'D\right] \cdot \pder{T}{t_n}~,\(rec)
$$
which in components reads
$$
\pmatrix{\pder{\tau}{t_{n+1}}\cr
\pder{u}{t_{n+1}}\cr} =
\pmatrix{\kappa^2\partial^2 + 2 u + 2 \partial^{-1} u \partial &
2\partial\tau\partial^{-1} - 2 \partial^{-1}\tau\partial\cr
0&\kappa^2\partial^2 + 2 u + 2\partial u\partial^{-1}\cr}\cdot
\pmatrix{\pder{\tau}{t_n}\cr
\pder{u}{t_n}\cr}~,\()
$$
and this, in turn, agrees with the recursion relation (40) in
\[Beckers].  Thus, we conclude that the flows of the two hierarchies
agree.

\section{Conserved Charges}

As it is well known, the conserved charges of the KdV hierarchy are
given by the traces of the fractional powers of the Lax operator,
namely (up to normalization)
$$
H_n = \Tr L^{n-1/2} = \int h_n\qquad\hbox{for
$n=1,2,\ldots$}~,\()
$$
where $h_n$ is the residue of $L^{n-1/2}$.  For the SKdV-B hierarchy,
the relevant Lax operator is $L = \kappa^2 \d^2 + T'$.  For such a Lax
operator, $h_n$ is a superfield, whence $H_n$ still has $\theta$
dependence:
$$
H_n[u,\tau] = A_n[u,\tau] + \theta B_n[u,\tau]~.\(charges)
$$
Notice that since $T'|_{\theta=0} = u$, $A_n$ is simply the $n^\th$
conserved charge $H_n^{\rm KdV}$ of the KdV hierarchy---in particular,
it is independent of $\tau$.  It is clear that both $A_n$ and $B_n$
are conserved under the SKdV-B flows, but in order to consider them as
conserved charges of a supersymmetric hierarchy, one has to take into
account the requirement that they be invariant under supersymmetry.
Making use of the above remark, let us rewrite \(charges) as follows:
$$
\int h_n(T') = \int h_n(u) + \theta \int b_n(u,\tau)~.\(chargestoo)
$$
Because $h_n(T')$ is a differential polynomial in $T'$, it transforms
under supersymmetry as a superfield.  In other words,
$$
\delta h_n(u) = b_n(u,\tau)\quad{\rm and}\quad \delta b_n(u,\tau) =
h_n(u)'~,\()
$$
whence $\delta B_n[u,\tau] = \int \delta b_n(u,\tau) = \int h_n(u)' =
0$.  That is, the charges $B_n$ are both supersymmetric and conserved.
Because they are supersymmetric they can be written as integrals over
superspace.  In fact,
$$
B_n[u,\tau] = \delta \int h_n(u) = \int \vder{h}{u}\cdot \tau' = \int
Dh_n\left(T'\right)|_{\theta=0} = \sint h_n(T')\,\()
$$
by definition of the Berezin integral.  It remains to show that these
charges are nontrivial.  Now, it is a classic result \[EarlyKdV] that
the $h_n(u)$ have the form $h_n(u) \propto u^n + \cdots$, where
$\cdots$ stand for terms with derivatives and hence a smaller number
of $u$'s.  Therefore,
$$
\vder{B_n}{T} \propto n(n-1) (T')^{n-2} T'' + \cdots\,\()
$$
which is nonzero for $n\geq 2$.  For $n=1$, $A_1 = \int u = \sint T$
is already supersymmetric.  In summary, we have proven the following
result.

\Prop<Two>{The conserved charges of the SKdV-B hierarchy are given by
$$
\eqnalign{H_n^{\rm SKdV-B} &= \delta H_n^{\rm KdV}\quad\hbox{for $n\geq
2$},\cr
           H_1^{\rm SKdV-B} &= H_1^{\rm KdV}~,\cr}
$$
and they obey the following relation (for $n\geq 2$)
$$
\Tr L^{n-1/2} = H_n^{\rm KdV}[u]+\theta H_n^{\rm
SKdV-B}[u,\tau]~,\(dubl)
$$
where $L = \kappa^2 \d^2 + T'$.\QED}

One can nevertheless ask the question whether these are in fact all
the conserved charges of the SKdV-B hierarchy.  We have found out by
hand one ``exotic'' charge
$$
H^{\rm exotic}=\int_B TT'~,\()
$$
and we have verified that there are no other exotic charges up to
weight  $\fr15/2$, where we say that $T$ has weight $\fr3/2$ and $D$
has weight $\fr1/2$.

\section{SKdV-B as a Reduction of SKP-type Hierarchies}

Since the SKdV-B flows are given by the isospectral
deformations of the Lax operator $L = \kappa^2 \d^2 + T'$, it is easy
to see that SKdV-B is but a particular reduction of the $\hbox{SKP}_2$
hierarchy introduced in \[SKPtwo].  First of all it is clear that $L$
has a unique square root of the form
$$
L^{1/2} = \kappa \d + \sum_{i\geq 1} A_i(T') \d^{1-i}~,\(sqrt)
$$
where the $A_i(T')$ are $\d$-differential polynomials in $T'$.
In terms of $L^{1/2}$, the flows defining SKdV-B are given by
$$
\pder{L^{1/2}}{t_n} \propto \comm[L^{n-1/2}_+,L^{1/2}]~.\(skdvbflows)
$$
Now, $L^{1/2}$ is but a special case of the general $\hbox{SKP}_2$
operator $\Lambda = \kappa \d + \sum_{k\geq1} B_k(T) D^{2-k}$
treated ($\kappa$ aside) in \[SKPtwo], where the $B_k(T)$ are
$D$-differential polynomials in $T$.  Moreover the $\hbox{SKP}_2$
flows are given by $\pder{\Lambda}{t'_n} =
\comm[\Lambda^n_+,\Lambda]$, which agree (after relabeling and
rescaling the times) with \(skdvbflows).  In other words, the
submanifold of $\hbox{SKP}_2$ operators of the form \(sqrt) is
preserved by the $\hbox{SKP}_2$ flows and, moreover, these flows agree
with the ones defining SKdV-B.

Moreover, since the Lax operator $L = \kappa^2 \d^2 + T'$ can be
``undressed'', one can map the SKdV-B hierarchy into the even part of
the SKP hierarchy of \[MaRa] or, equivalently, the Jacobian SKP
hierarchy of \[MuRa].  To this effect, let us define an element $S$ of
the Volterra group by $L = S \kappa^2 \d^2 S^{-1}$.  In terms of $S$,
the SKdV-B flows can be written as (up to $\kappa$ factors)
$$
\pder{S}{t_n} \propto - (S \d^{2n+1} S^{-1})_- S~.\(dressing)
$$
This equation is then the one defining the even flows of the SKP
hierarchy, when we think of $S$ as an element of the larger
superVolterra group.

\section{Bihamiltonian Structure and Integrability}

It was shown in \[SKPtwo] that SKdV-type reductions of the
$\hbox{SKP}_2$ hierarchy are bihamiltonian: the two structures being
given by the supersymmetric analogs of the Gel'fand--Dickey brackets
constructed in \[slax].  In particular, the hierarchy associated to
the operator $D^4 + U_1D^3 + U_2D^2 + U_3D + U_4$ is bihamiltonian,
and so is its reduction $U_1=U_2=U_4=0$ to SKdV.  It would thus seem
reasonable to expect that the SKdV-B hierarchy, which is obtained as
the reduction $U_1=U_2=U_3=0$ and $U_4=T'$, would inherit a
bihamiltonian structure in this fashion.  However, this turns out not
to be the case: it is easy to show that setting $U_1=U_2=U_3=0$
collapses the rest of the phase space.

We can nevertheless exhibit a bihamiltonian structure for SKdV-B
exploiting its equivalence with KdV($T'$).  We first rewrite the
analogs of \(gdone) and \(gdtwo) for KdV($T'$):
$$
\veqnalign{\pder{T'}{t_k} &= \d \cdot \left.\vder{H_{k+1}^{\rm
KdV}}{u}\right|_{u=T'}\(gdtone)\cr
    &= \left[ \kappa^2 \d^3 + 2 T' \d + 2 \d T'\right] \cdot
\left.\vder{H_k^{\rm KdV}}{u}\right|_{u=T'}~.\(gdttwo)\cr}
$$
For $H_k^{\rm KdV} = \int h_k(u)$, we have that
$$
\veqnalign{\left.\vder{H_k^{\rm KdV}}{u}\right|_{u=T'} &=
\sum_{i\geq0} (\d^i)^*\cdot \left.\pder{h_k}{u^{(i)}}\right|_{u=T'}
\cr
&= \sum_{i\geq0} (D^{2i})^*\cdot \pder{h_k}{T^{[2i+1]}}\cr
&= - D^{-1} \sum_{i\geq0} (D^{2i+1})^*\cdot
\pder{h_k}{T^{[2i+1]}}\cr}
$$
Since $h_k(T')$ only depends on the odd $D$-derivatives of $T$ we may
add for free the contribution of the even derivatives, and we obtain
\fnote{It should be noticed that the above formula does {\it not} mix
grading.  This follows from the definition of the variational
derivative in the formal calculus of variations, in which integration
is simply the operation of dropping total derivatives.  This point
seems to have caused some confusion in the literature \[Confused].}
$$
\left.\vder{H_k^{\rm KdV}}{u}\right|_{u=T'} = -D^{-1}\cdot
\sum_{i\geq 0} (D^i)^*\pder{h_k}{T^{[i]}} = D^{-1} \cdot
\vder{H_k^{\rm SKdV-B}}{T}\,\()
$$
for $H_k^{\rm SKdV-B} = \sint h_k(T')$.  We can thus rewrite
\(gdtone) and \(gdttwo) as follows
$$
\veqnalign{\pder{T}{t_k} &= \vder{H_{k+1}^{\rm
SKdV-B}}{T}\()\cr
    &= \left[ \kappa^2 \d^2 + 2 D^{-1} T' D + 2 D T' D^{-1}\right]
\cdot \vder{H_k^{\rm SKdV-B}}{T}~.\cr}
$$

These equations look already to be in hamiltonian form, with Poisson
structures $J_1 = 1$ and $J_2 =  \kappa^2 \d^2 + 2 D^{-1} T' D + 2 D
T' D^{-1}$.  Notice that $J_1$ satisfies the Jacobi identities
trivially, since it is constant.  It may seem at first odd that it is
not antisymmetric---but this is nothing new in supersymmetric
hierarchies, which can have both even and odd Poisson structures
\[Oddbracks].  The second structure $J_2$ may not seem obviously
Poisson, but it is not hard to show that the Jacobi identities are
satisfied.  Notice that $J_2$ also defines odd Poisson brackets which
are moreover nonlocal.  This is again nothing new in supersymmetric
hierarchies: the first Poisson structure of SKdV is also nonlocal;
although the flows, just like the ones here, are local.  Notice,
parenthetically, that as expected $J_2\, J_1^{-1}$ coincides with the
recursion operator \(rec) for SKdV-B.

Finally, notice that $J_1$ can be obtained from $J_2$ by shifting $T'
\mapsto T' + \lambda$.  Since $J_2$ is Poisson for any $T$, it follows
that $J_1$ and $J_2$ are coordinated.  Usual arguments now imply that
the conserved charges are in involution relative to both Poisson
structures.  In summary, SKdV-B is an integrable bihamiltonian
supersymmetric hierarchy.

\section{Some Remarks on Odd flows}

Although as proven in \<FlowsComm> the odd flows are not
supersymmetric, it is possible to modify them in such a way that they
are.  First of all notice that the explicit expression \(tau) of
$\tau$ as a function of the odd times and the $R_k$ can only be
reconciled with its transformation law \(SUSY) under supersymmetry, if
$\tau_1$ transforms under supersymmetry.  To see this, let us plug
\(tau) into the second equation of \(SUSY):
$$
\veqnalign{u &= - \sum_{k\geq 0} (\delta \tau_k) R_k + \sum_{k\geq 0}
\tau_k \delta R_k&\cr
&= - \sum_{k\geq 0} (\delta \tau_k) R_k - \sum_{k\geq 0}
\tau_k \pder{}{t_{n-1}} \sum_{\ell\geq 0} \tau_{\ell}
R_{\ell}&\hbox{by \(flowsontau) and \(tau)}\cr
&= - \sum_{k\geq 0} (\delta \tau_k) R_k - \sum_{k,\ell\geq 0}
\tau_k \tau_{\ell} \pder{R_{\ell}}{t_{n-1}}\cr
&= - \sum_{k\geq 0} (\delta \tau_k) R_k&\hbox{by \(flowscommute)}\cr}
$$
which implies that
$$
\delta\tau_k = - \delta_{k,1}~.\(deltatauone)
$$
Consider now the flows given by
$$
D_n \equiv \pder{}{\tau_n} - \tau_1 \pder{}{t_{n-1}}~.\()
$$
{}From \(deltatauone) and \(oddflows) it follows that these flows are
supersymmetric.  It is moreover obvious that they commute with the
even flows, and that all $D_{n\neq1}$ (anti)commute among themselves.
The remaining algebra of flows is
$$
D_1^2 = - \d\quad\hbox{and}\quad\comm[D_1,D_n] = -\pder{}{t_{n-1}}
\quad \forall n>1~,\()
$$
where we have used the fact that $\pder{}{t_0} = \d$.  This defines a
supersymmetric extension of the SKdV-B hierarchy by odd flows.

It now remains to find a representation of the above algebra of flows
in superspace.  The main obstacle lies in that the $D_n$ explicitly
depend on $\tau_1$ which, as \(deltatauone) suggests, should be
represented as $-\theta$.  It is easy to check that the representation
induced from $\delta \mapsto Q$ and $\tau_1 \mapsto -\theta$ is
inconsistent, and we have thus far been unable to find a consistent
superspace representation for the odd flows.

Alternatively one could try to induce odd flows via the embedding of
the SKdV-B hierarchy into the even part of (a reduction of) the SKP
hierarchy hierarchy of \[MuRa].  The even flows of both the SKP
hierarchy of \[MaRa] and the Jacobian SKP hierarchy of \[MuRa] agree,
but the odd ones don't.  Nevertheless, via \(dressing), we can
understand these flows as flows in the superVolterra group.  It is
easy to see that the flows of neither of the two hierarchies preserve
the Volterra subgroup where $S$ lives.

\section{Epilogue}

We can thus understand the SKdV-B hierarchy as a (perhaps somewhat
naive) supersymmetrization of the KdV hierarchy.  It thus behooves us
to ask whether other reductions of the KP hierarchy may be
supersymmetrized in this fashion.  It turns out that of the
generalized KdV hierarchies, only the Boussinesque admits this
supersymmetrization.\fnote{Of course, the naive supersymmetrization of
substituting the bosonic fields by even superfields which are not
derivatives of anything can always be achieved, but these don't seem
to be the ones that appear in the supersymmetric matrix models.}
Brevity demands that we omit the details, which will appear somewhere
else \[Bouss].  As a closing remark, let us simply add that if, as
expected, the supersymmetrized Boussinesque hierarchy alluded to above
plays a role in the double scaling limit of the supersymmetric analog
of the two-matrix model, then something interesting should happen in
the supersymmetric three-matrix model.

\ack

We are grateful to K.~Becker and M.~Becker for their interesting
correspondence, and to E.~Ramos for his insightful and stimulating
conversations as well as for a careful reading of the manuscript.  In
addition, S.S.~is grateful to W.~Nahm and V.~Rittenberg for their kind
support and encouragement.
\refsout
\bye